\documentclass[aps,prl,superscriptaddress,twocolumn,showpacs,]{revtex4}

\usepackage{graphicx}
\usepackage{dcolumn}
\usepackage{bm}
\usepackage{amsfonts}
\usepackage{amsmath}
\usepackage{color}
\usepackage{braket}
\usepackage{tikz}
\usepackage{ulem}

\begin{document}

\title{The tunneling density-of-states of interacting massless Dirac fermions}
\author{A. Principi}
\affiliation{NEST, Istituto Nanoscienze-CNR and Scuola Normale Superiore, I-56126 Pisa, Italy}
\author{Marco Polini}
\email{m.polini@sns.it}
\affiliation{NEST, Istituto Nanoscienze-CNR and Scuola Normale Superiore, I-56126 Pisa, Italy}
\author{Reza Asgari}
\affiliation{School of Physics, Institute for Research in Fundamental Sciences (IPM), Tehran 19395-5531, Iran}
\author{A.H. MacDonald}
\affiliation{Department of Physics, University of Texas at Austin,
Austin, TX, 78712, USA}

\begin{abstract}
We calculate the tunneling density-of-states (DOS) of a disorder-free two-dimensional interacting electron system 
with a massless-Dirac band Hamiltonian. 
The DOS exhibits two main features: i) linear growth at large energies with a slope that is suppressed by quasiparticle velocity enhancement, 
and ii) a rich structure of plasmaron peaks which appear at negative bias voltages in an n-doped sample  
and at positive bias voltages in a p-doped sample. 
We predict that the DOS at the Dirac point is non-zero even in the absence of 
disorder because of electron-electron interactions,
and that it is then accurately proportional to the Fermi energy.
The finite background DOS observed 
at the Dirac point of graphene sheets and topological insulator surfaces can therefore be  
an interaction effect rather than a disorder effect.  
\end{abstract}

\maketitle

\noindent
{\it Introduction:}  A {\it V}-shaped density-of-states (DOS) curve, $\nu_0(E) \propto |E|/v^2$ where $E$
is energy measured from the Dirac point, is a distinctive feature of graphene and of topologial insulator (TI) 
surface states.  Both two-dimensional electron systems are described by free-particle 
massless Dirac fermion (MDF) Hamiltonians:
$h_0({\bm k}) = \hbar v {\bm \sigma} \cdot {\bm k}$~\cite{graphenetheoreticalreviews,TIreviews}. 
Here ${\bm \sigma} = (\sigma^x,\sigma^y)$ is a 2D vector of Pauli matrices and
$v$ is the MDF velocity, which is roughly three-hundred times smaller than the speed of light in
the graphene sheet case~\cite{graphenetheoreticalreviews,grapheneexperimentalreviews}, and six-hundred times smaller~\cite{TIreviews} 
in the case of the currently-studied chalcogenide TI's.

The DOS of electronic systems has often been measured using tunneling, and 
in recent decades often using scanning tunneling microscopy (STM) which probes locally to mitigate the
influence of inhomogeneities and some types of disorder.
In STM~\cite{sts_review} the DOS is proportional to the differential conductivity $dI_{\rm tip}/dV_{\rm bias}$,
where $I_{\rm tip}$ and $V_{\rm bias}$ are respectively the tunneling current and
the bias voltage between the STM tip and the sample.
When graphene sheets~\cite{STMgraphene} or TI surfaces~\cite{STMTI} are 
studied by STM spectroscopy,
the experimental DOS curves are never simple {\it V}-shaped curves. 
This is unsurprising since disorder, electron-phonon, and electron-electron interactions~\cite{manoharangroup} can all play a role  
in altering the tunneling DOS. 
Brar {\it et al.}~\cite{brar_prl_2010} have however recently discovered 
some very specific gate-voltage dependent, {\it i.e.} carrier density dependent, features in the $dI_{\rm tip}/dV_{\rm bias}$ spectra of doped graphene sheets deposited on ${\rm SiO}_2$ which they attributed to electron-plasmon interactions. 
Similar features with an enhanced amplitude have been observed by the same group
for graphene sheets deposited on h-BN~\cite{decker_nano_2011}, suggesting that the effect is rather universal.

In this Rapid Communication we propose electron-electron interactions as the source of 
these DOS spectral features.  Indeed theory has 
already predicted~\cite{polini_prb_2008,hwang_prb_2008} that electron-plasmon interactions have a large impact on the one-body spectral function ${\cal A}(k,\omega)$ of doped graphene.  After carrying out extensive studies of 
${\cal A}(k,\omega)$ in quasi-freestanding doped graphene sheets on (hydrogen-terminated) SiC using angle-resolved photoemission spectroscopy (ARPES), Bostwick {\it et al.}~\cite{bostwick_science_2010} recently discovered experimentally
that coupling between electrons and plasmons leads to a substantial reconstruction of the MDF conical spectrum.
The reconstruction is related to the appearance of new composite quasiparticles, known as plasmarons~\cite{earlywork}, 
which are expected to appear when interactions between electrons and the collective charge-density (plasmon) oscillations 
of an electron liquid~\cite{Giuliani_and_Vignale} are strong enough.
Plasmarons are signaled theoretically by multiple solutions of the one-particle Dyson equation, and 
experimentally by satellite bands in the ARPES spectra. 
Since the tunneling DOS is given by an integral over all momenta of ${\cal A}(k,\omega)$, 
plasmaron-related features must also be present in the STM spectra.

\begin{figure}[t]
\begin{center}
\includegraphics[width=0.95\linewidth]{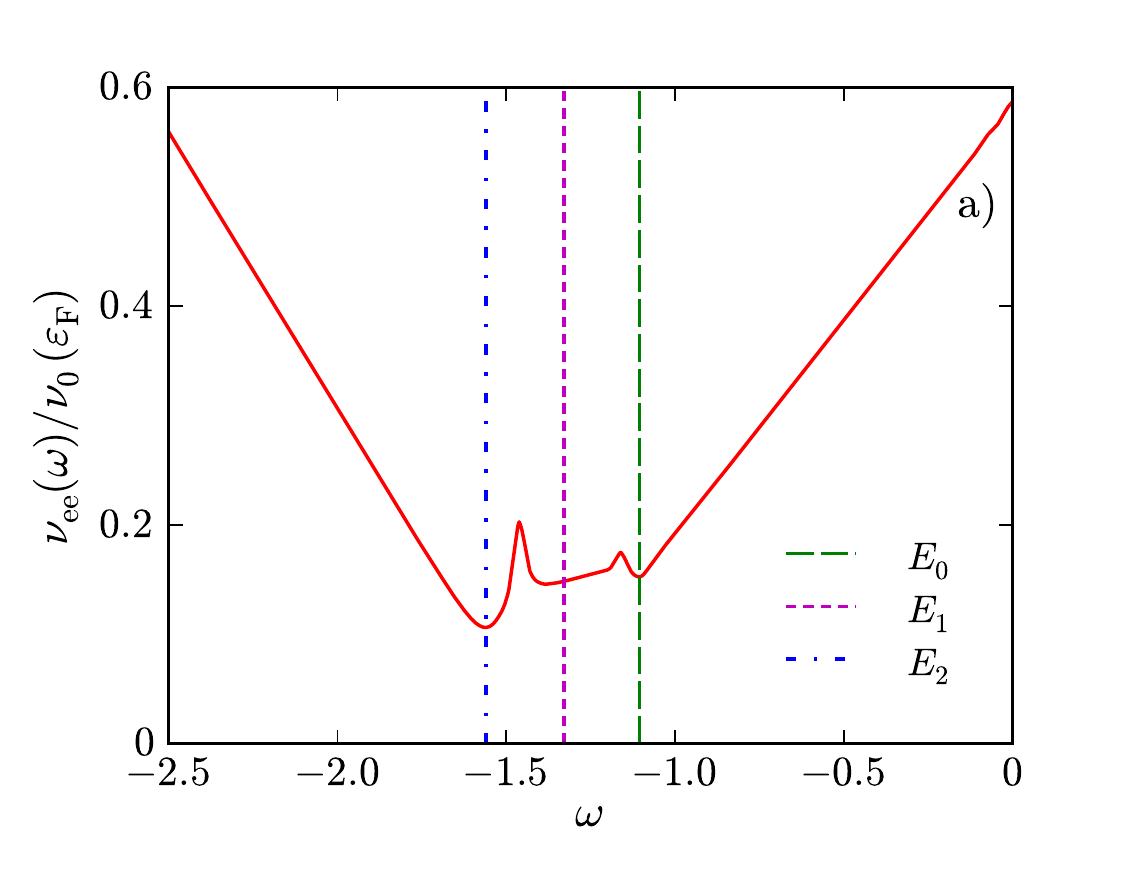}\\
\includegraphics[width=0.9\linewidth]{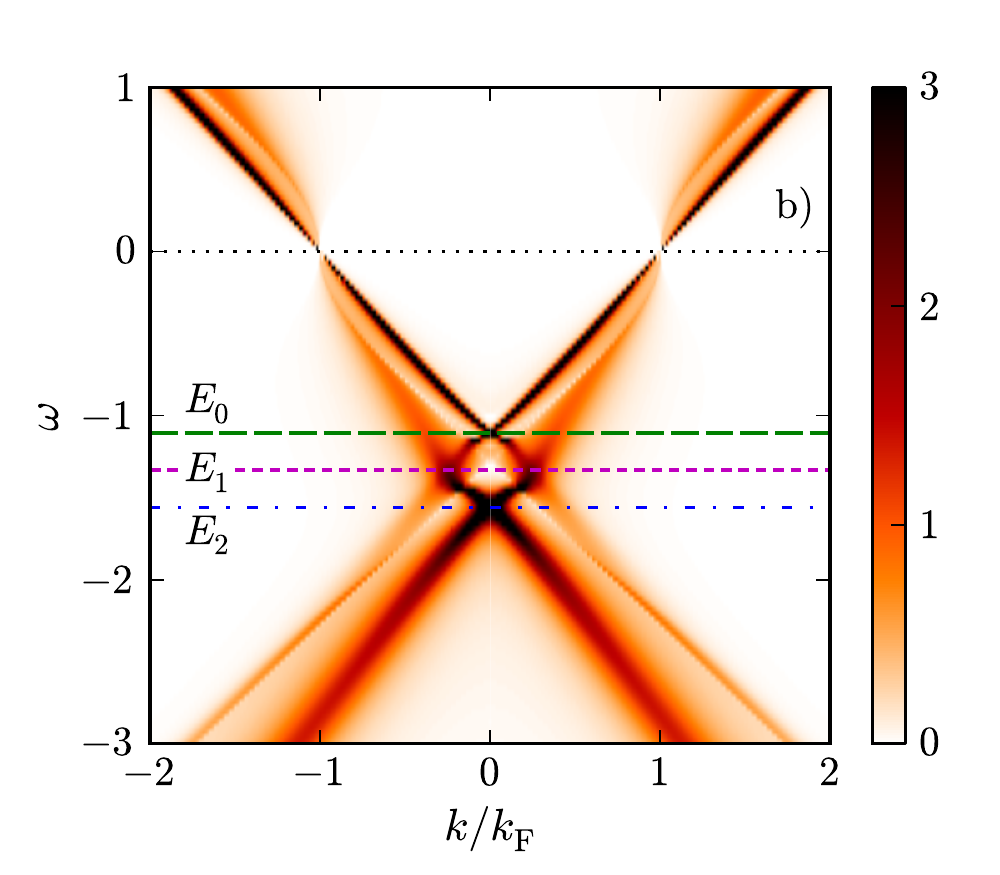}
\caption{(Color online) Panel a) Tunneling density-of-states (DOS) $\nu_{\rm ee}(\omega)$
of electrons with a MDF band Hamiltonian as a function of energy $\omega$ 
measured from the chemical potential $\mu$.  The DOS is nearly independent of carrier density when it is  
expressed in units of the non-interacting DOS at the Fermi energy $\nu_0(\varepsilon_{\rm F})$,
and $\omega$ is expressed in units of $\varepsilon_{\rm F}$. 
These results are for $N_{\rm f}=4$, $n=10^{12}~{\rm cm}^{-2}$, and $\alpha_{\rm ee} = 0.5$. 
Panel b) A 2D color plot of the spectral function ${\cal A}(k,\omega)$ for the same parameters used in panel a).
The energies $E_0$ (long-dashed line), $E_1$ (short-dashed line), and $E_2$ (dash-dotted line) define 
a diamond-like shape in $(k,\omega)$ space~\cite{bostwick_science_2010} in which the
influence of electron-plasmon coupling is strongest.
The DOS has measurable spectral structure in the energy range $E_2 < \omega < E_0$.
\label{fig:one}}
\end{center}
\end{figure}

Fig.~\ref{fig:one}a) summarizes our theoretical results for the tunneling DOS of 
electrons with MDF bands which were obtained 
using a continuum model $G_0W$ (random phase)
approximation~\cite{polini_prb_2008,hwang_prb_2008},
with the dimensionless Coulomb coupling constant $\alpha_{\rm ee} = e^2/(\hbar v \epsilon)=0.5$ 
and the flavor number $N_{\rm f}=4$  set to values appropriate for graphene on h-BN. 
The flavor number accounts for graphene's spin and valley degeneracies.  ($N_{\rm f}=1$
for TI surfaces.)  The DOS differs from the non-interacting result 
$\nu_0(E) = N_{\rm f} |E|/(2\pi \hbar^2 v^2)$, in three ways:
i) at large energies the DOS {\it is} linear, but its slope is altered because the 
quasiparticle velocity 
$v^\star$ is renormalized~\cite{elias_natphys_2011,borghi_ssc_2009}; 
ii) the DOS does not vanish at any energy; and 
iii) the DOS displays a rich structure of plasmaron peaks separated by local minima located near the energies
of the Dirac crossings~\cite{bostwick_science_2010} in the corresponding ARPES spectrum [Fig.~\ref{fig:one}b)].

In what follows we introduce and discuss qualitative features of
the continuum model Hamiltonian and 
the random phase ($G_0W$) approximation that we have employed in combination
to approximate 2D MDF interaction physics to obtain the results illustrated in Fig.~\ref{fig:one}. 
We then compare the roles of electron-electron 
and electron-disorder interaction in filling in the Dirac model DOS {\it V},
concluding that interaction effects play the dominant role in high quality graphene samples.

\noindent
{\it The Coulomb MDF model:} 
We describe 2D electron systems with  MDF bands 
using the following Hamiltonian (setting $\hbar \to 1$ from now on):
\begin{equation} \label{eq:MDFhamiltonian}
{\hat {\cal H}} =  v\sum_{{\bm k}, \alpha, \beta} {\hat \psi}^\dagger_{{\bm k}, \alpha} 
( {\bm \sigma}_{\alpha\beta} \cdot {\bm k} ) {\hat \psi}_{{\bm k}, \beta}
+
\frac{1}{2 S}\sum_{{\bm q} \neq {\bm 0}} v_q{\hat \rho}_{\bm q} {\hat \rho}_{-{\bm q}}
~.
\end{equation}
Here $v$ is the bare Fermi velocity , $S$ is the sample area, and ${\hat \psi}^\dagger_{{\bm k}, \alpha}$ (${\hat \psi}_{{\bm k}, \alpha}$) creates (destroys) an electron with momentum ${\bm k}$ and two-valued (pseudo)spin index $\alpha$,
and flavor indices are implicit.
In the case of graphene, $\alpha= A,B$ is a sublattice-pseudospin index, whereas in the 
case of TI surface states it is a true spin index.  In Eq.~(\ref{eq:MDFhamiltonian})
$
{\hat \rho}_{\bm q} = \sum_{{\bm k}, \alpha} {\hat \psi}^\dagger_{{\bm k} - {\bm q}, \alpha}{\hat \psi}_{{\bm k}, \alpha}
$
is the density operator, and
$
v_q =2\pi e^2/(\epsilon q)
$
is the 2D Fourier transform of Coulomb potential. Here $\epsilon$ depends on the dielectric environment surrounding the 2D MDF fluid. In the case of graphene $\epsilon=(\epsilon_1 + \epsilon_2)/2$, where $\epsilon_1$ ($\epsilon_2$) is the dielectric constant of the medium above (below) the sheet.  For the surface states of a thick TI $\epsilon=(\epsilon_{1}+\epsilon_{\rm TI})/2$
where $\epsilon_{\rm TI}$ is the TI bulk dielectric constant.  Because chalcogenide TI's have large dielectric constants, 
interaction effects on their properties will tend to be weak.  For this reason we concentrate on graphene MDF's in what follows.
Interaction effects can be made stronger in chalcogenide TI MDF's by preparing thin film samples, but interactions between
top and bottom surfaces will then play an essential role.  

Below we describe an electron-doped system with an excess electron density $n$; 
the properties of hole-doped systems can be obtained by appealing to the model's particle hole symmetry 
as we explain below.  The corresponding Fermi wave number and Fermi energy are $k_{\rm F} = \sqrt{4\pi n/N_{\rm f}}$ and $\varepsilon_{\rm F}=v k_{\rm F}$, respectively.  In calculating the DOS it will be convenient to use the 
Fermi level rather than the Dirac point as the zero of energy so that 
the Dirac-band single-particle energies
are $\xi_{{\bm k},\lambda} \equiv \varepsilon_{{\bm k}, \lambda} - \varepsilon_{\rm F} = \lambda v k - \varepsilon_{\rm F}$
where $\lambda=\pm 1$ distinguishes the conduction and valence bands.
Note that the band and interaction terms scale in the same way with 
changes in length scale.  The spectra we calculate will therefore be independent of 
$k_{\rm F}$ if energies are measured in terms of $\varepsilon_{\rm F}$, 
apart from weak logarithmic dependences on a $\pi$-band width 
ultraviolet cutoff scale which must be included in the theory. 
It follows in particular that the DOS at the Dirac point is accurately proportional to the 
Fermi energy.

\noindent
{\it $G_0W$ approximation spectral functions:} The DOS is related to the spectral function ${\cal A}(k, \omega)$ by
\begin{eqnarray} \label{eq:DOS}
\nu_{\rm ee}(\omega) &=& N_{\rm f} \int \frac{d^2{\bm k}}{(2\pi)^2} \; {\cal A}(k, \omega) \nonumber\\
&=& \int_{0}^{+\infty}dE~ \; \nu_0(E) \, {\cal A}(E/v,\omega)~,
\end{eqnarray}
where ${\cal A}(k,\omega) = \sum_{\lambda = \pm}{\cal A}_\lambda(k, \omega)$ and (suppressing the $k,\omega$ variables for simplicity)
\begin{equation} \label{eq:A_lambda}
{\cal A}_\lambda = \frac{1}{\pi}\frac{|\Im m~\Sigma_{\lambda}|}{(\omega - \xi_{{\bm k}, \lambda} - \Re e~\Sigma_{\lambda})^2+ (\Im m~\Sigma_{\lambda})^2}~.
\end{equation}
The quantity $\Sigma_{\lambda}(k, \omega)$ denotes the electron-electron interaction
contribution to the quasiparticle self-energy. The real part of $\Sigma_{\lambda}(k, \omega)$ is measured from its value at the Fermi surface, which physically represents the interaction contribution to the chemical potential, $\mu_{\rm int}$. Correspondingly, the variable $\omega$ in Eqs.~(\ref{eq:DOS})-(\ref{eq:A_lambda}) [and also in Eq.~(\ref{eq:Sigma_res}) below] represents energy measured from the chemical potential $\mu = \varepsilon_{\rm F} + \mu_{\rm int}$ of the interacting system and in units of the Fermi energy $\varepsilon_{\rm F}$. 
Physically, ${\cal A}_\lambda(k, \omega)$ represents~\cite{Giuliani_and_Vignale} the probability density for increasing or decreasing the energy of the $N$-particle system by $\omega$ upon adding or removing a single particle in state $|{\bm k}, \lambda\rangle$. An STM experiment probes both occupied ($\omega< 0$) and empty ($\omega>0$) states, whereas ARPES probes the occupied portion of the spectrum only.

For the quasiparticle self-energy we use the $G_0W$ (or random phase)
approximation~\cite{Giuliani_and_Vignale,polini_prb_2008,hwang_prb_2008,bostwick_science_2010}, in which the self-energy is expanded up to first order in the dynamically-screened interaction $W_{\rm ee}(q,\omega) = v_q / \varepsilon_{\rm RPA}(q,\omega)$: 
\begin{eqnarray} \label{eq:Sigma_res}
\Im m~\Sigma_{\lambda}(k, \omega) &=&
\sum_{\lambda'} \int \frac{d^2 {\bm q}}{(2\pi)^2}~\Im m\left[W_{\rm ee}(q,\Omega_{{\bm k}^+,\lambda'})\right]
\nonumber\\
&\times&
{\cal F}_{{\bm k}\lambda,{\bm k}^+\lambda'}
\left[\Theta(\Omega_{{\bm k}^+,\lambda'})-\Theta(-\xi_{{\bm k}^+,\lambda'}) \right]
~,
\nonumber\\
\end{eqnarray}
where $\Omega_{{\bm k},\lambda} = \omega-\xi_{{\bm k}, \lambda}$, ${\cal F}_{{\bm k}\lambda,{\bm k}'\lambda'} = [1+\lambda\lambda'\cos(\varphi_{\bm k}-\varphi_{{\bm k}'})]/2$ is the MDF model chirality factor, and ${\bm k}^+$ is a shorthand for ${\bm k}+{\bm q}$. The RPA dielectric function is 
$
\varepsilon_{\rm RPA}(q,\omega) = 1 - v_{q} \chi^{(0)}_{\rho\rho}(q,\omega)
$, 
where $\chi^{(0)}_{\rho\rho}(q,\omega)$ is the well-known Lindhard response function of non-interacting 2D MDFs~\cite{wunsch_njp_2006} at arbitrary doping $n$. The real part of the self-energy is obtained from Eq.~(\ref{eq:Sigma_res}) by a Kramers-Kronig transform.

Note that, since the MDF model is particle-hole symmetric, the real and imaginary
[$\Re e~\Sigma_{\lambda}(k,\omega)$ and $\Im m~\Sigma_{\lambda}(k,\omega)$]
parts of the quasiparticle self-energy for electron doping 
are equal to  $-\Re e~\Sigma_{-\lambda}(k,-\omega)$
and $\Im m~\Sigma_{-\lambda}(k,-\omega)$ for hole doping. 
Thus ${\cal A}_\lambda (k, \omega)$ and $\nu_{\rm ee}(\omega)$ for electron doping 
are equal to ${\cal A}_{-\lambda} (k, -\omega)$ and $\nu_{\rm ee}(-\omega)$ for hole doping.

\noindent
{\it Interaction Strength and Disorder Dependence:} As seen in Fig.~\ref{fig:one}b), the plasmaron bands loose spectral weight as they approach the 
Fermi energy but are otherwise remarkably well defined, providing a lower energy echo of the 
quasiparticle Dirac point. The diamond-like shape in $(k,\omega)$ space~\cite{bostwick_science_2010} that can be clearly recognized in Fig.~\ref{fig:one}b)
is bounded by plasmaron and quasiparticle peaks which cross at a finite value of $k$ at an energy $E_1$ 
which lies between the quasiparticle band crossing at energy $E_0$ and the plasmaron 
band crossing at energy $E_2$.  
\begin{figure}[t]
\begin{center}
\begin{tabular}{c}
\includegraphics[width=0.95\linewidth]{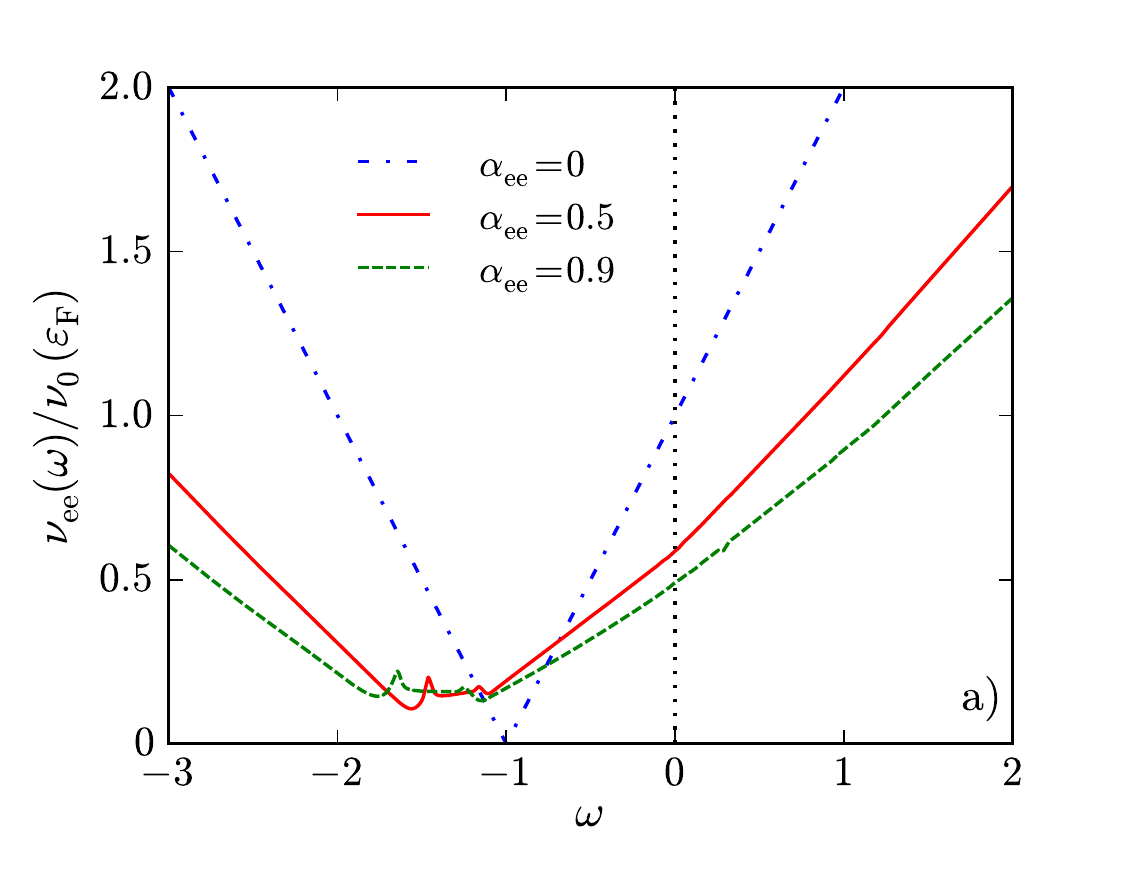} \\
\includegraphics[width=0.95\linewidth]{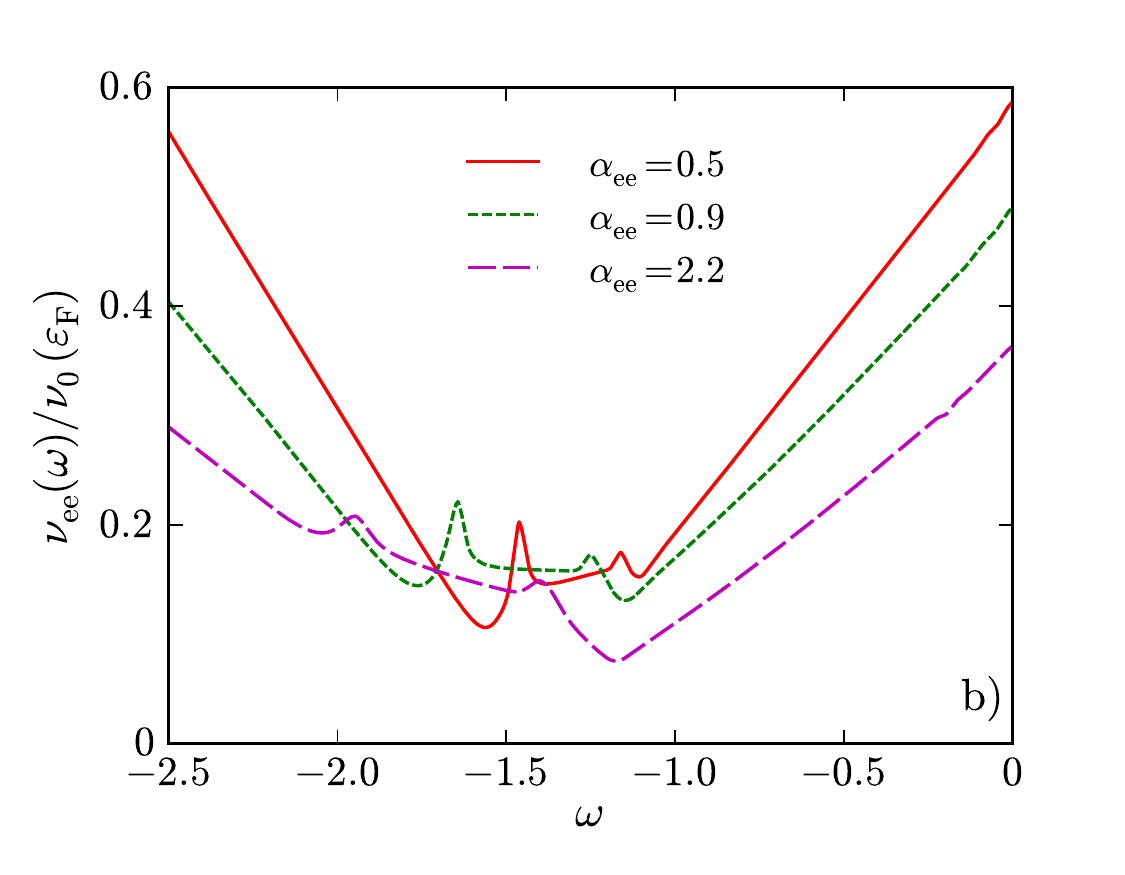}
\end{tabular}
\caption{(Color online)
Panel a) DOS of a 2D interacting MDF system [in units of the non-interacting DOS at the Fermi energy $\nu_0(\varepsilon_{\rm F})$] as a function of energy $\omega$ (measured from the chemical potential $\mu$ and in units of the Fermi energy $\varepsilon_{\rm F}$). The results plotted in 
this figure are for $N_{\rm f} = 4$, carrier density $n=10^{12}~{\rm cm}^{-2}$, and for three values of the coupling constant $\alpha_{\rm ee}$ (including the non-interacting case $\alpha_{\rm ee} = 0$). The vertical dotted line represents the chemical potential $\mu$ of the interacting system. Panel b) Expanded view of the energy region near the reconstructed Dirac point.  
The features shown here originate from $(k,\omega)$ diamond explained in the text and 
illustrated in Fig.~\ref{fig:one}b). In this panel we have also included data for $\alpha_{\rm ee} = 2.2$,
a value that is suitable  to describe a {\it suspended} graphene sheet. 
Note that the diamond features in $\nu_{\rm ee}(\omega)$ 
are strongest for $\alpha_{\rm ee}  \sim 0.5$, a value thought to be appropriate to a graphene sheet on h-BN,
and not for the strongest interactions considered.\label{fig:two}}
\end{center}
\end{figure}
The diamond feature represents a massive reconstruction of the MDF spectrum.
As shown in Figs.~\ref{fig:one}a) and~\ref{fig:two}, it has a striking impact on the DOS at low energies. 
Note that for a n-doped system the strongest plasmaron features appear in the filled portion of the spectra.  The particle-hole symmetry properties explained above imply that 
the opposite occurs for a p-doped system. At large energies we clearly see that the DOS grows linearly with a slope
which decreases as $\alpha_{\rm ee}$ increases, an effect which can be understood 
qualitatively in terms of the interaction enhancement~\cite{borghi_ssc_2009}
of the quasiparticle Fermi velocity $v^\star$.

\begin{figure}[t]
\begin{center}
\includegraphics[width=0.95\linewidth]{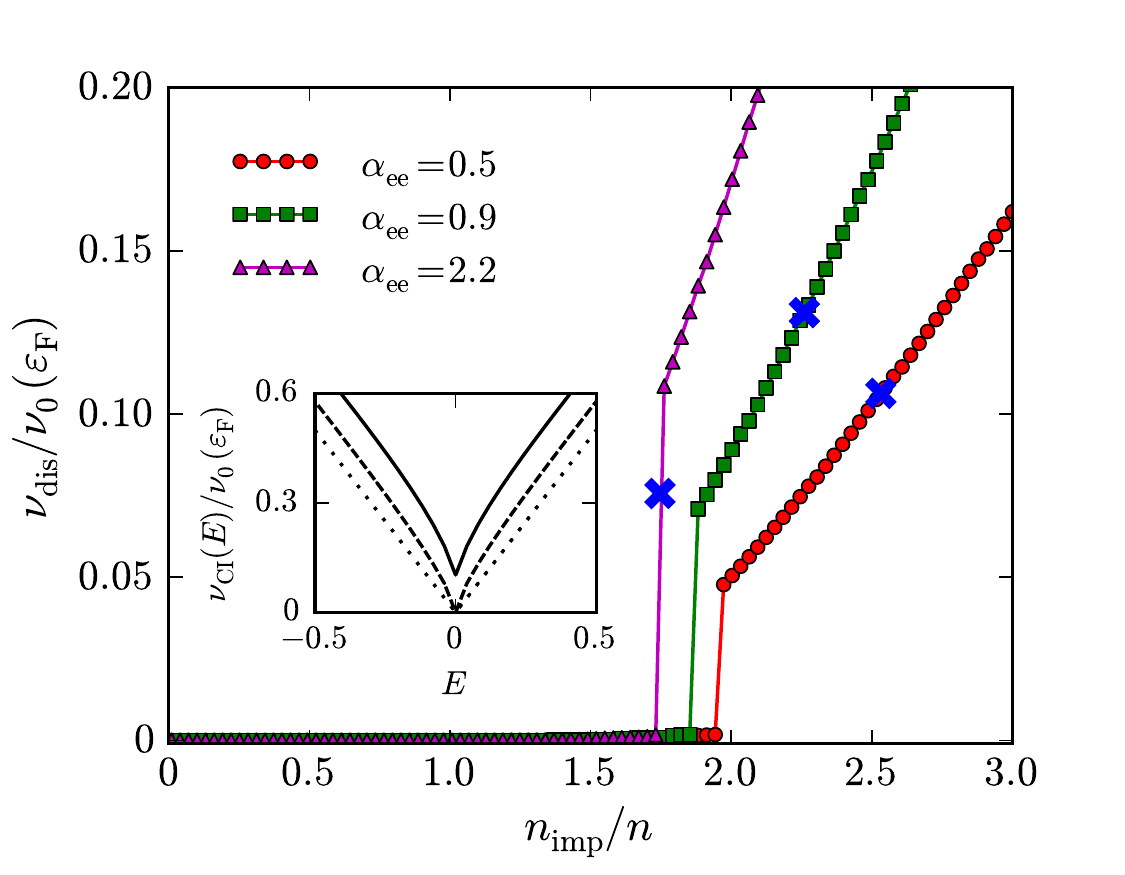}
\caption{(Color online) DOS of a 2D non-interacting MDF system at the Dirac-point energy in the presence of Coulomb scatterers with $Z =1$ located on the 2D plane where electrons move [$d=0$ in Eq.~(\ref{eq:SE_SCBA})].  The dimensionless parameter on the horizontal axis is the impurity concentration, {\it i.e.} $n_{\rm imp}$ in units of the electron density $n$.  The DOS disorder background $\nu_{\rm dis} = \nu_{\rm CI}(0)$ is plotted in units of the non-interacting MDF clean-system DOS at the Fermi energy, {\it i.e.} $\nu_0(\varepsilon_{\rm F})$, to facilitate comparison with Fig.~\ref{fig:two}. 
The three sets of data refer to three values of the coupling constant $\alpha_{\rm ee}$. The solid lines are guides to the eye.
The thick (blue) crosses indicate the values of impurity concentration at which $\nu_{\rm dis}$ equals $\nu_{\rm bg}$,  the background DOS of the clean interacting system at the same value of $\alpha_{\rm ee}$. 
The two contributions are clearly comparable only for impurity concentrations well above unity. The inset shows $\nu_{\rm CI}(E)$ as a function of energy $E$ measured from the Dirac point (and in units of the Fermi energy). Results in this figure are based on the SCBA - Eq.~(\ref{eq:SE_SCBA}).  The abruptness of the DOS increase is certainly an artifact of the SCBA.
\label{fig:three}}
\end{center}
\end{figure}

In Fig.~\ref{fig:two} we plot the DOS of an interacting 2D MDF system for three values of the coupling constant $\alpha_{\rm ee}$. Due to the increase of screening with increasing $\alpha_{\rm ee}$, the amplitude of the plasmaron peaks in the energy range $E_2 < \omega < E_0$ becomes quite small when $\alpha_{\rm ee} \gtrsim 2$. 
The DOS does {\it not} vanish at any energy when interactions are 
included, as observed experimentally in Ref.~\onlinecite{brar_prl_2010}.
The strength of the interaction induced background in the non-interacting {\it V} 
can be characterized by the absolute minimum value of 
$\nu_{\rm ee}(\omega)$: $\nu_{\rm bg} = \min_{\omega}[\nu_{\rm ee}(\omega)]$.
To estimate how important this background typically is for the interpretation of experiment, we 
can compare it with the finite DOS background due to disorder.  To this end, we calculate the DOS of a 2D MDF system in the presence of a model disorder potential, 
including electron-electron interactions only through the static RPA screening function $\varepsilon_{\rm RPA}(q,0)$. 
For the sake of definiteness, we consider Coulomb impurities (CI) located at random in a plane parallel to the 2D electron system
that is separated by distance $d$. Let $n_{\rm imp}$ be the impurity density.
For our qualitative goals it is sufficient to use the self-consistent Born approximation (SCBA)~\cite{vozmediano_prb_2010}
for which the disorder self-energy reads
\begin{equation} \label{eq:SE_SCBA}
\Sigma^{({\rm CI})}_\lambda(k, E) = \sum_{\lambda'} \int\frac{d^2{\bm q}}{(2\pi)^2}
\frac{n_{\rm imp}W^2_{\rm ei}(q,0){\cal F}_{{\bm k}\lambda,{\bm q}\lambda'}}{E - \varepsilon_{{\bm q}, \lambda'} -\Sigma^{({\rm CI})}_{\lambda'}(q, E)}
~,
\end{equation}
where $W_{\rm ei}(q,0) = v_{\rm ei}(q)/\varepsilon_{\rm RPA}(q,0)$ is the statically-screened electron-impurity potential,
and $v_{\rm ei}(q,0) = v_q Z \exp(-qd)$ is the 2D Fourier transform of the bare electron-impurity potential (assuming that all impurities have charge equal to $Ze$, in absolute value). The DOS in the presence of disorder, $\nu_{\rm CI}(E)$, can be computed by iteratively evaluating the integral in Eq.~(\ref{eq:SE_SCBA}) numerically until $\Sigma^{({\rm CI})}_\lambda(k, E)$ is self-consistent.

In Fig.~\ref{fig:three} we plot the DOS background due to long-range scatterers, $\nu_{\rm dis} = \nu_{\rm CI}(0)$, as a function of the impurity concentration $n_{\rm imp}/n$ and for different values of the fine-structure constant $\alpha_{\rm ee}$. Note, indeed, that since we have assumed Coulomb disorder, the strength of the electron-impurity interaction is controlled by the same dimensionless coupling constant which determines the strength of electron-electron interactions. We clearly see from this plot that the impurity concentration needed to give a background comparable to $\nu_{\rm bg}$ is rather high. For example, for graphene on ${\rm SiO}_2$ the impurity density $n_{\rm imp}$ necessary to make disorder more important than interactions is more than twice the electron density. 
Typical impurity densities, extracted from a comparison between theory and transport data~\cite{dassarma_rmp_2011} which
assumes Coulomb scatterers, are
for comparison on the order of $\approx 10^{10} - 10^{12} ~{\rm cm}^{-2}$, 
giving typical values of $n_{\rm imp}/n \lesssim 1$. 
We thus conclude that the background due to electron-electron interactions is very significant, at least in comparison with that due to long-range Coulomb impurities.  Of course, the disorder effect will always dominate at very 
small carrier densities.  

In summary, we have calculated the tunneling DOS of a 2D system of interacting MDFs. 
We have found that plasmarons in the one-body spectral function give rise to a non-trivial structure of peaks and
valleys in the DOS. The spectral weight carried by plasmarons gives rise to a finite background in the DOS, which 
does not vanish at any energy and instead displays two peaks associated with the  
electron and plasmaron Dirac points~\cite{bostwick_science_2010}. 
A more direct comparison with the experimental results 
in Ref.~\onlinecite{decker_nano_2011}  
is necessary to determine convincingly 
whether or not the DOS peak has already been observed in the STM spectra of graphene on h-BN. 
Our predictions can also be tested by low-temperature STM spectroscopy on quasi-freestanding doped graphene sheets on (hydrogen-terminated) SiC~\cite{NEST}.
Finally, our findings are also relevant to STM studies of TI surface states.
A finite background has, for example, recently been observed by Alpichshev {\it et al.}~\cite{STMTI} in the STM spectra of the surface states of ${\rm Bi}_2{\rm Te}_3$ which could be due in part to interactions.

{\it Acknowledgements. ---} Work in Pisa was supported by the Italian Ministry of Education, University, and Research (MIUR) through the program ``FIRB - Futuro in Ricerca 2010" (project title ``PLASMOGRAPH: plasmons and terahertz devices in graphene").  A.H.M. was supported by the Welch Foundation under grant TBF1473 and by the NRI SWAN program.
As this manuscript was being completed we became aware of recent closely related work by LeBlanc {\it et al.}~\cite{leblanc_prb_2011}. 
The authors of this work also predict distinct plasmaron features in the DOS of an interacting system of MDFs. 


\begin{thebibliography}{77}
%
\bibitem{graphenetheoreticalreviews}
	A.H. Castro Neto {\it et al.}, \rmp {\bf 81}, 109 (2009).
%
\bibitem{TIreviews}
	M.Z. Hasan and C.L. Kane, \rmp {\bf 82}, 3045 (2010);
	J.E. Moore, Nature {\bf 464}, 194 (2010); 
	X.-L. Qi and S.-C. Zhang, \rmp {\bf 83}, 1057 (2011).
%
\bibitem{grapheneexperimentalreviews}
	A.K. Geim and K.S. Novoselov, Nature Mater. {\bf 6}, 183 (2007);
	A.K. Geim, Science {\bf 324}, 1530 (2009).
%
\bibitem{sts_review}
	J. Tersoff and D.R. Hamann, \prb {\bf 31}, 805 (1985); 
	C.J. Chen, {\it Introduction to Scanning Tunneling Microscopy} (Oxford University Press, New York, 2008).
%
\bibitem{STMgraphene}	
	For a recent review see {\it e.g.} M. Morgenstern, Phys. Status Solidi (b) {\bf 248}, 2423 (2011).
%
\bibitem{STMTI}
	See {\it e.g.} P. Roushan {\it et al.}, Nature {\bf 460}, 1106 (2009);
	T. Zhang {\it et al.}, \prl {\bf 103}, 266803 (2009);
	Z. Alpichshev {\it et al.}, \prl {\bf 104}, 016401 (2010);
	Y. Okada {\it et al.}, {\it ibid.} {\bf 106}, 206805 (2011);
	H. Beidenkopf {\it et al.}, Nature Phys. advance online publication, 09 October 2011 (DOI 10.1038/nphys2108).
%
\bibitem{manoharangroup}
	To the best of our knowledge, the only 2D MDF systems to date that display an almost perfect 
	{\it V}-shaped DOS are graphene flakes on graphite [G. Li, A. Luican, and E.Y. Andrei, \prl {\bf 102}, 176804 (2009)] 
	and {\it artificial graphene} lattices created by placing 
	atoms one-by-one on the surface of a metal with a honeycomb lattice arrangement 
	[K.K. Gomes {\it et al.}, http://meetings.aps.org/link/BAPS.2011.MAR.A21.8].
%
\bibitem{brar_prl_2010}
	 V.W. Brar {\it et al.}, \prl {\bf 104}, 036805 (2010).
%
\bibitem{decker_nano_2011}
	R. Decker {\it et al.}, Nano Lett. {\bf 11}, 2291 (2011).
%
\bibitem{polini_prb_2008}
	M. Polini {\it et al.}, \prb {\bf 77}, 081411(R) (2008); 
	see also A. Principi, R. Asgari, and M. Polini, Solid State Commun. {\bf 151}, 1627 (2011).
%
\bibitem{hwang_prb_2008}
	E.H. Hwang and S. Das Sarma, \prb {\bf 77}, 081412(R) (2008).	
%
\bibitem{bostwick_science_2010}
	A. Bostwick {\it et al.}, Science {\bf 328}, 999 (2010); 
	A.L. Walter {\it et al.}, \prb {\bf 84}, 085410 (2011).
%
\bibitem{earlywork}
	B.I. Lundqvist, Phys. Kondens. Mater. {\bf 6}, 193 (1967);
	L. Hedin {\it et al.}, Solid State Commun. {\bf 5}, 237 (1967);
	for recent experimental work on plasmarons see 
	R. Tediosi {\it et al.}, \prl {\bf 99}, 016406 (2007) and O.E. Dial {\it et al.}, arXiv:1009.0519.
%
\bibitem{Giuliani_and_Vignale}
	G.F. Giuliani and G. Vignale, {\it Quantum Theory of the Electron Liquid} (Cambridge University Press, Cambridge, 2005).
%
\bibitem{elias_natphys_2011}
	D.C. Elias {\it et al.}, Nature Phys. {\bf 7}, 701 (2011).
%
\bibitem{borghi_ssc_2009}
	J. Gonz\'alez, F. Guinea, and M.A.H. Vozmediano, Nucl. Phys. B {\bf 424}, 595 (1994) and 
	\prb {\bf 59}, R2474 (1999); G. Borghi {\it et al.}, Solid State Commun. {\bf 149}, 1117 (2009).
%
\bibitem{wunsch_njp_2006}
	See {\it e.g.} B. Wunsch {\it et al.}, New J. Phys. {\bf 8}, 318 (2006).
%
\bibitem{vozmediano_prb_2010}
	F. de Juan, E.H. Hwang, and M.A.H. Vozmediano, \prb {\bf 82}, 245418 (2010).
%
\bibitem{dassarma_rmp_2011}
	S. Das Sarma {\it et al.}, \rmp {\bf 83}, 407 (2011).
%
\bibitem{NEST}
	S. Goler {\it et al.}, submitted.
%
\bibitem{leblanc_prb_2011}
	J.P.F. LeBlanc, J.P. Carbotte, and E.J. Nicol, \prb {\bf 84}, 165448 (2011).
%
\end{thebibliography}
\end{document}